# Control of superluminal transit through a heterogeneous medium

M. Kulkarni and N. Seshadri

St. Stephen's College, Delhi 110007, INDIA

and

V. S. C. Manga Rao and S. Dutta Gupta*

School of Physics, University of Hyderabad, Hyderabad 500046, INDIA

Email: sdgsp@uohyd.ernet.in

ABSTRACT

We consider pulse propagation through a two component composite medium (metal inclusions in a dielectric host) with or without cavity mirrors. We show that a very thin slab of such a medium, under conditions of localized plasmon resonance, can lead to significant superluminality with detectable levels of the transmitted pulse. A cavity containing the heterogeneous medium is shown to lead to subluminal-to-superluminal transmission depending on the volume fraction of the metal inclusions. The predictions of the phase time calculations are verified by explicit calculations of the transmitted pulse shapes. We also demonstrate the independence of phase time on system width and the volume fraction under specific conditions.

* Author for correspondence

## I. Introduction

In recent years there has been a great deal of interest in subluminal and superluminal propagation of pulses through dispersive media [1,2]. A clever manipulation of the dispersive properties of the medium has made ‘slow’ light a reality [3]. On the other hand, since the pioneering experiment of Chu and Wong [4], there have been several reports (both theoretical and experimental) of superluminal transit in a variety of systems [5-16]. In some of these studies [13-15] the key issue has been the saturation exhibited by the pulse advancement as a function of system width, referred to as the Hartman effect [17]. The origin and the manifestations of the Hartman effect (independence of phase time on the system width) were recently investigated by Winful [14]. Generally speaking, the superluminal transit takes place in presence of evanescent waves in the structure. Note that the structure then acts as a typical ‘barrier’ in an equivalent problem of tunnelling of a quantum particle. Thus optical stratified media can be the prototypes for one dimensional quantum scattering problems. The delay/advancement through the optical system can be examined by looking at the phase time [18], which is given by the frequency derivative of the phase of the transmission coefficient.

It is now clear that dispersion management is the key component leading to the subluminal or superluminal propagation of pulses. From this viewpoint, the scope of heterogeneous media becomes clear immediately. Two- or multi-component heterogeneous media (both linear and non-linear) have attracted a lot of attention for their potential for diverse applications [19, 20]. The linear properties of a two component composite material can be well represented by Maxwell-Garnett theory, where the volume fraction of the inclusion plays a very important role. In case of metal inclusions (with negative real part of the dielectric function), localized plasmon resonance, leading to enhanced local fields, can be excited. The strength, location and width of the localized plasmon resonance depend crucially on the volume fraction. Thus, at the preparation stage the volume fraction can be used as an effective handle for controlling the dispersive and absorptive properties of the heterogeneous material. In this paper, we study the transmission characteristics of such a medium enclosed in a Fabry-Perot cavity (made of silver mirrors). The motivation of introducing the cavity will be clear if one recalls the opposite nature of dispersion of an empty cavity and the bare localized plasmon resonance. In the vicinity of the resonance the cavity offers normal dispersion [2] leading to subluminal propagation. In contrast, the bare plasmon resonance much like a resonant absorber offers anomalous dispersion causing superluminal transit [4]. Thus one expects an interesting interplay of these opposing effects leading to an effective control of the pulse delay/advancement. We show that one can indeed have a substantial range of pulse velocities, from subluminal to superluminal. Moreover, we demonstrate normal mode splitting, reminiscent of vacuum field Rabi splitting [21], when the cavity is tuned to the localized plasmon resonance. Finally we study the delay saturation effects in a slab of the composite medium and show the independence of the phase time on the width of the slab, as well as on the volume fraction, under specific conditions.

The organisation of the paper is as follows. In Section II, we present the model and give the outline of the method. In Section III, we present the results and discuss them. Finally in Section IV, we summarize the main results of the paper.

## II. Mathematical Formulation

Consider the system shown in figure 1, where a slab of the heterogeneous medium of length $d$ is enclosed between two silver mirrors, each of thickness $d_1$. The heterogeneous medium consists of silver inclusions (with dielectric function $\varepsilon_1(\omega)$), embedded in a dielectric host (with dielectric constant $\varepsilon_h$). We assume the host material to be dispersionless, while for silver dielectric function, we use the interpolated experimental data of Johnson and Christy [22]. Under suitable conditions, the optical properties of the intra-cavity medium can be approximated by the effective dielectric function $\overline{\varepsilon}(\omega)$ given by

$$\overline{\varepsilon}(\omega) = \varepsilon_h + \frac{fx(\varepsilon_1 - \varepsilon_h)}{1 + f(x-1)}; \quad x = \frac{3\varepsilon_h}{\varepsilon_1 + 2\varepsilon_h}, \qquad (1)$$

where $f$ is the volume fraction. In writing Eq.(1), we assumed the linear dimension of the inclusions to be small enough (compared to the wavelength $\lambda$), for quasistatics to be applicable; at the same time, the grains should be large enough so that they may be represented by bulk properties. It is clear from Eq.(1) that for metals ($\mathrm{Re}(\varepsilon_1)$<0), there can be resonant enhancement for $\mathrm{Re}(\varepsilon_1) + 2\varepsilon_h = 0$. We will show that such resonances, often referred to as localized plasmon resonances, play a key role in manipulating the pulse characteristics. The scope of these resonances in the context of delay control and superluminality will be clear if one looks at the wavelength dependence of the effective refractive index $\overline{n}(\omega) = \sqrt{\overline{\varepsilon}(\omega)}$ and relates it to the group index $n_g$ (see figure 2) given by

$$n_g = \bar{n}_r + \omega \frac{\partial \bar{n}_r}{\partial \omega}, \tag{2}$$

where $\bar{n}_r$ is the real part of $\bar{n}(\omega)$. Recall that a group index smaller (larger) than unity leads to superluminal (subluminal) transit. It is clear from figure 2 that there is a wavelength region where the frequency derivative of $\bar{n}_r$ is negative with the magnitude of $\bar{n}_r$ less than unity leading to a highly negative group index. Thus, a suitable choice of the pulse carrier frequency in this region can lead to significant superluminality. The major problem with superluminal transit is the attenuation of the pulse amplitude. Since, in our case, one can choose a carrier slightly away from the centre of the absorption band, the effects of pulse attenuation can be made to be less significant.

We now assume that a pulse of the form

$$E(z,t) = A(\bar{t})e^{-i\omega_c \bar{t}}\ ;\ \bar{t} = t - \frac{z}{c}, \tag{3}$$

is normally incident on the system. Let the envelope of the pulse be given by

$$A(t) = \exp[-(t/\tau)^2], \tag{4}$$

where $2\tau$ is the temporal width of the pulse. In terms of the input given by Eqs.(3) and (4), the transmitted pulse can be written as

$$E_T(z,t) = \int_{-\infty}^{\infty} d\omega\, A(\omega)\, T(\omega) e^{-i\omega(t-\frac{z-d_t}{c})}, \tag{5}$$

where $d_t = d + 2d_1$ is the total length of the system, $A(\omega)$ is the Fourier image of $A(t)$ and $T(\omega)$ is the complex amplitude of the transmission coefficient through the structure. The amplitude transmission for the layered medium at any frequency $\omega$ can be calculated using the standard characteristic matrix approach [23]. It is then straight forward to calculate the phase time by taking the derivative of the phase of the transmission coefficient [18] as follows

$$\tau_p = \left(\frac{\partial \Phi}{\partial \omega}\right)_{\omega=\omega_c}, \tag{6}$$

where $\Phi$ is the phase of the complex transmission coefficient. It is clear that the phase time (6) needs to be compared with the free space transit time $\tau_f = d_t / c$ in order to infer about the delay or advancement. The calculations based on Eq.(6) can be checked against the explicit output and input pulse profiles given by Eqs.(5) and (4), respectively. In the next Section we present the numerical results pertaining to the transmission coefficient, phase times and incident and transmitted pulse shapes.

## III. Numerical results and discussions

A very preliminary study of the wavelength dependence of the effective index of the composite medium in the last section revealed the scope of such media for probing the superluminal propagation of pulses. In order to demonstrate this, we consider the system shown in figure 1 with the following parameter set: $d_1 = 0.025\mu$m, $d = 1.4257\mu$m, $\varepsilon_h = 2.25$. As mentioned earlier, we used the experimental data of Johnson and Christy [22] for obtaining (by spline interpolation) the dielectric function of silver. The width of the cavity was chosen such that one of the empty-cavity (filled only with host, i.e., $f = 0.0$) resonance is close to the localized plasmon resonance for the bulk heterogeneous medium with finite volume fraction. Recall that the localized plasmon resonance gets red shifted with an increase in volume fraction $f$. It is clear that the width of the silver mirrors determines their reflectivity and hence the quality factor of the cavity. The width was controlled in such a way that the cavity resonance width without silver inclusions ($f = 0.0$) is comparable to that of the localized plasmon resonance. The motive is straightforward – to make the oscillators (the cavity and the localized plasmon modes) close enough so that one can realize the conditions of normal mode splitting. The situation is quite similar to the vacuum field Rabi splitting [21], where the cavity modes are strongly coupled to the intra-cavity atoms, except for a crucial difference. A change in the number of atoms in the cavity leads to a change in the coupling strength only, leaving the location and the width of the resonance intact. In our case, the analogue of the density is the volume fraction of the inclusions and an increase in volume fraction leads to a red shift, as well as a broadening of the localized plasmon resonance. Nevertheless, sufficient coupling (i.e. larger volume fraction) can lead to a splitting of the otherwise degenerate modes. This splitting is shown in figure 3a, where we have plotted the absolute value of the amplitude transmission $T$ through the cavity as a function of wavelength $\lambda$. The curves from top to bottom (e.g., at $\lambda \sim 0.405\mu$m) are for $f = 0.00$, 0.0001, 0.001 and 0.004, respectively. For reference, in the same plot we have shown the localized plasmon resonance ($\mathrm{Im}(\bar{\varepsilon})$) for $f = 0.004$ by a dashed line. The results for phase time are shown in figure 3b, where we have plotted the difference of phase time $\tau_p$ and free space transit time $\tau_f$ as a function of wavelength. The curves from top to bottom are for the same set of values of $f$ as in figure3a. It is clear from figure 3b that without the inclusions in the intra-cavity medium, the pulse transmission (close to resonance) is subluminal. With an increase in the volume fraction, the dynamics changes over from subluminal to superluminal. Thus for larger values of $f$ one expects highly superluminal transit albeit at the cost of higher attenuation of the pulse. In fact one can have significant superluminality still with detectable levels of transmitted pulse. Some such cases are shown in figure 4 for different values of $f$ (see curves in the lower panel). We have compared the transmitted pulse profile with that of the reference pulse (incident pulse shifted by $\tau_f$). For convenience of comparison, the input pulse amplitude (unity) has been scaled down by the peak value of the transmitted pulse amplitude. All the results of figure 4 are for a pulse with $\tau = 1.0\,\mathrm{ps}$ centred about the carrier wavelength $\lambda_c = 0.40465\mu\mathrm{m}$. The delay/advancement for all values of $f$ is in excellent agreement with the predictions of the phase time calculations (see figure 3b). Note that the advancement can be quite significant (see figure 4 for $f$=0.004) amounting to a ratio $(\tau_p - \tau_f)/\tau_f \sim -27.35$ with 0.082% peak amplitude transmission. In units of pulse width ($2\tau$) the advancement is about -6.7%. For $f$=0.001, the same are given by -6.94, 9.12% and -1.7%, respectively. Note also that the last figure can be improved considerably for shorter pulses since the phase time is not affected much by the width of the pulse. At this stage a few comments regarding the distortion of the pulse in propagation would be in place. It may be noted here that for $\tau = 1.0$ ps, except for the inevitable attenuation, there are no significant distortions in the pulse shape. This is due to the fact that over the range of the pulse spectrum, the dispersive properties of the system do not change appreciably. This is not the case for narrower pulses, say, with $\tau = 0.1$ ps and 0.01ps (curves not shown). In these cases, especially for $\tau = 0.01$ ps, significant deviations (like stretching and ringing [16]) of the pulse profile from the gaussian input profile were noted.

We next look for Hartman effect for a slab of the heterogeneous medium (without mirrors). For $f = 0.1$ we pick a wavelength, say, $\lambda = 0.4$ μm, where the propagation is superluminal. Note that for the said parameter values, $\mathrm{Re}(\overline{\varepsilon}) < 0$ and the waves are evanescent in the slab. Thus, the slab acts as a 'barrier' and one expects a saturation of phase time with increasing slab width $d$. However, the calculated phase time for such a structure exhibits a negative slope as a function of $d$ (see curve 2 in figure 5a). This deviation from the usual Hartman effect can be easily explained if one ignores the cavity effect (due to finite reflection at the edges of the slab) and writes the phase as $\Phi = (\omega / c)\overline{n}_r d$. The derivative $D_d^{\tau}$ of the corresponding phase time with respect to the slab width $d$ can then be written as

$$D_d^{\tau} = \frac{\partial \tau_p}{\partial d} = \frac{1}{c}\left( \overline{n}_r + \omega \frac{\partial \overline{n}_r}{\partial \omega} \right). \qquad (7)$$

It is clear that Eq.(7) needs to be evaluated at the pulse carrier frequency. The inset to figure 5a shows the plot of $D_d^{\tau}$ for $f = 0.1$ as a function of $\lambda$. It is clear from the inset that the slope of the curve $\tau_p$ vs $d$ is negative at $\lambda = 0.4$ μm, and it can be zero at two distinct points, namely at $\lambda = 0.3911$ μm and 0.420705 μm. The plot of $\tau_p$ vs $d$ (calculated as per Eq.(6), retaining the cavity effects) for the latter two values of the wavelength are shown in figure 5a, demonstrating clearly the independence of phase time on the width of the structure. Note that these operating points correspond to infinite group velocities since the corresponding group index (see Eq.(2)) is zero. It is not surprising then why the phase time becomes independent of the width of the structure. One other interesting aspect is the robustness of the saturation for $\lambda = 0.3911$ μm, where a small change in the wavelength does not change the slope appreciably. In contrast a slight deviation from $\lambda = 0.420705$ μm, leads to a drastic change in the slope. In the context of the Hartman effect, it should be noted that the same may be recovered in a 'pure' barrier obtained by suppressing the imaginary part of the dielectric function (i.e., by setting $\mathrm{Im}(\overline{\varepsilon}) = 0.0$ with $\mathrm{Re}(\overline{\varepsilon}) < 0.0$ ). In such a case the propagation constant in the slab is purely imaginary and one ends up with the usual Hartman effect (see the dashed curve in figure 5a for $\lambda = 0.4$ μm). Despite the lack of independence of phase time on the barrier width for arbitrary wavelengths, there are other interesting saturation effects. One such effect is shown in figure 5b, where we have plotted the phase time as a function of the volume fraction $f$ for three different values of $\lambda$. The dips near $f = 0.12$ are due to the sensitivity of the plasmon resonance to the volume fraction. Recall that the resonance gets red shifted with an increase in the volume fraction. As can be seen from this figure, the phase time becomes practically independent of $f$ for $f > 0.25$. This can be easily explained if one recalls that at these wavelengths the heterogeneous material makes a transition from dielectric to metal-like as the volume fraction of the metal inclusions is increased, making the barrier increasingly ‘pure’.

## IV. Conclusions

In conclusion, we studied pulse propagation through a cavity containing a metal-dielectric heterogeneous medium near the localized plasmon resonance. The volume fraction of the heterogeneous medium is shown to play a very important role. We have demonstrated normal mode splitting when the cavity is tuned to the localized plasmon resonance. The rich (and controllable) dispersive properties of the heterogeneous medium are shown to lead to both advancement and delay of pulses without distortion. We show significant superluminal advancement with appreciable signal levels. We also report new phase time saturation effects as functions of width of the slab (for specific wavelengths) and the volume fraction of the heterogeneous medium.

## Figure Captions

Fig.1. Schematic view of the Fabry-Perot cavity containing a slab of heterogeneous medium of width $d$ between two silver mirrors of width $d_1$ each.

Fig.2. Real and imaginary parts of effective refractive index $\overline{n}_r$ (solid), $\overline{n}_i$ (dashed), respectively, and the group index $n_g$ (dashed-dot) as functions of wavelength $\lambda$, for a heterogeneous medium with host dielectric constant $\varepsilon_h = 2.25$ and volume fraction $f = 0.1$.

Fig.3. (a) Absolute value of the amplitude transmission *T* versus wavelength $\lambda$ for a Fabry-Perot cavity for different values of $f$ (solid lines). The curves from top to bottom are for $f = 0.0, 0.0001, 0.001$, and 0.004, respectively. The dashed line shows $\mathrm{Im}(\overline{\varepsilon})$ for $f = 0.004$. For all the curves, $d = 1.4257\mu$m, $d_1 = 0.025\mu$m and $\varepsilon_h = 2.25$; (b) The difference between the phase time $\tau_p$ and the vacuum transit time $\tau_f$ plotted against wavelength for the same values of $f$ as in (a).

Fig.4. Reference (solid line) and transmitted (dashed line) pulse profiles for a gaussian input pulse of unit peak amplitude with $\tau = 1.0$ps, with carrier wavelength $\lambda_c = 0.40465\mu$m. The different curves are labelled by their corresponding values of $f$.

Fig.5. Phase time $\tau_p$ versus (a) $d$ for a heterogeneous slab of width $d$ and volume fraction $f = 0.1$; (b) volume fraction $f$ for a slab of width $d = 0.1\mu$m. Curves labelled by 1,2,3 are for $\lambda =$ $0.3911\,\mu m$, $0.4\,\mu m$ and $0.402705\,\mu m$,

respectively. The dashed line in (a) is the plot for $\lambda = 0.4\,\mu m$, but with $\mathrm{Im}(\bar{\varepsilon}) = 0$.

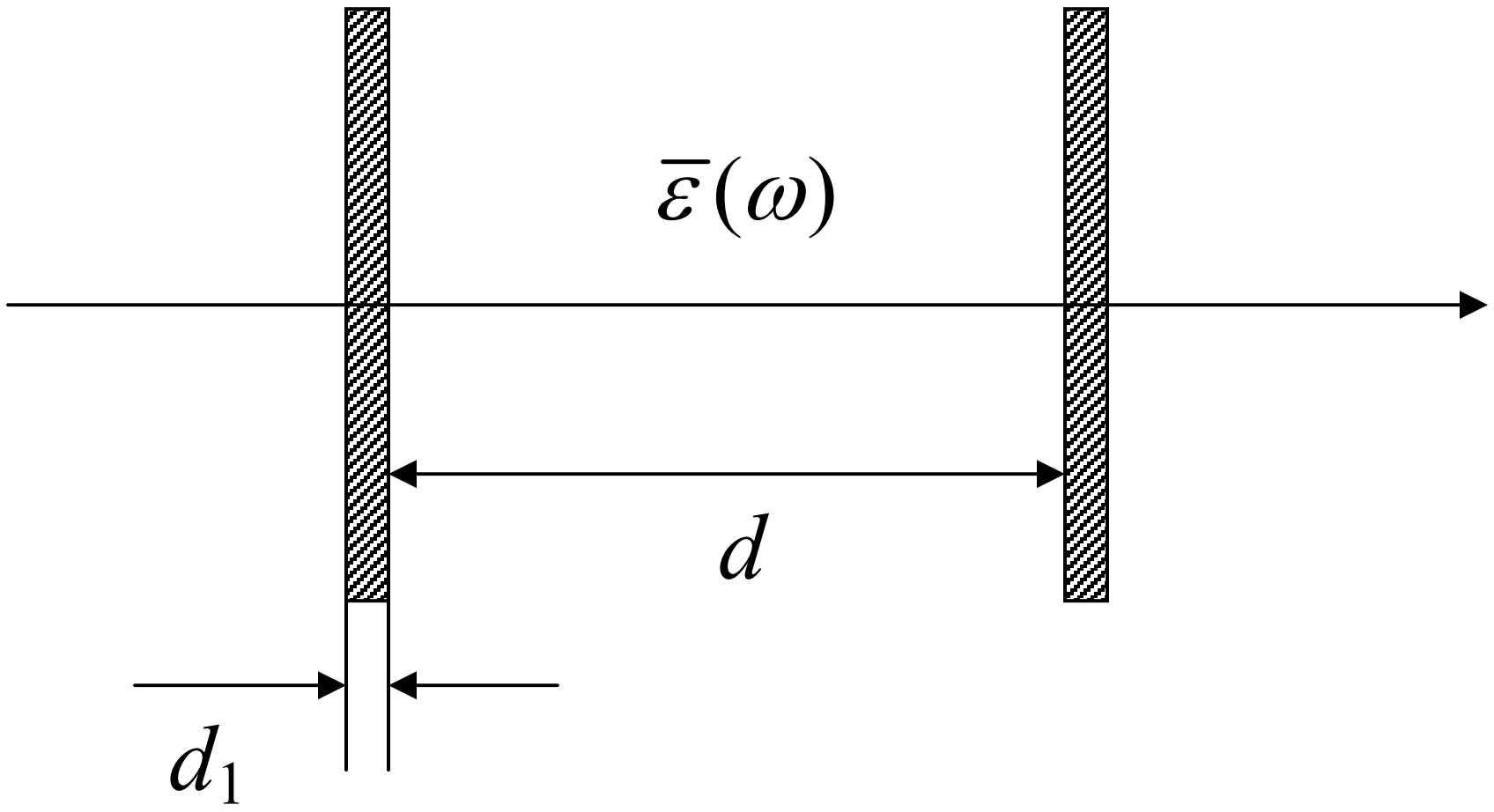

Fig.1 M Kulkarni..,"Control of superluminal …."

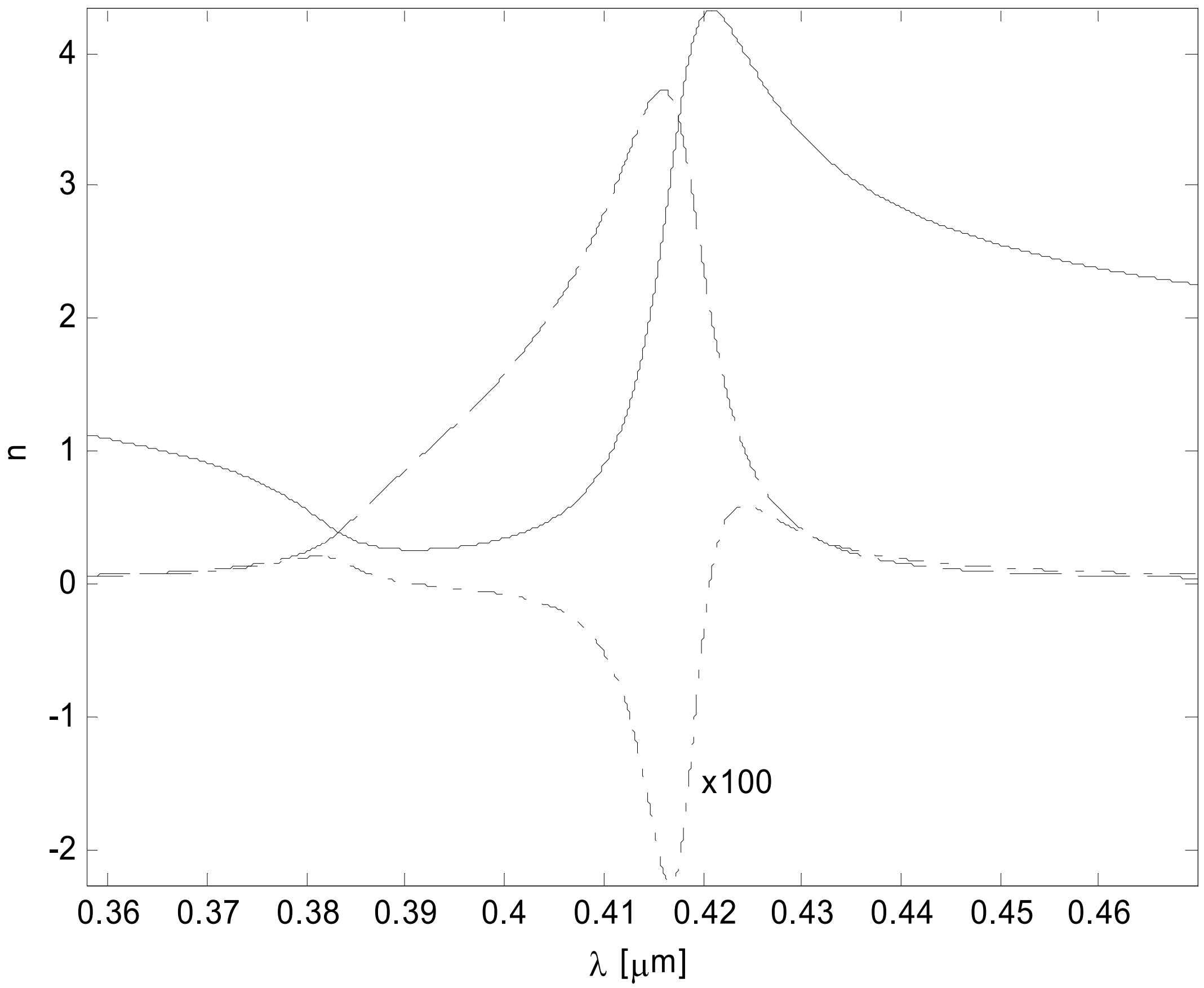

Fig.2 M Kulkarni..,"Control of superluminal …."

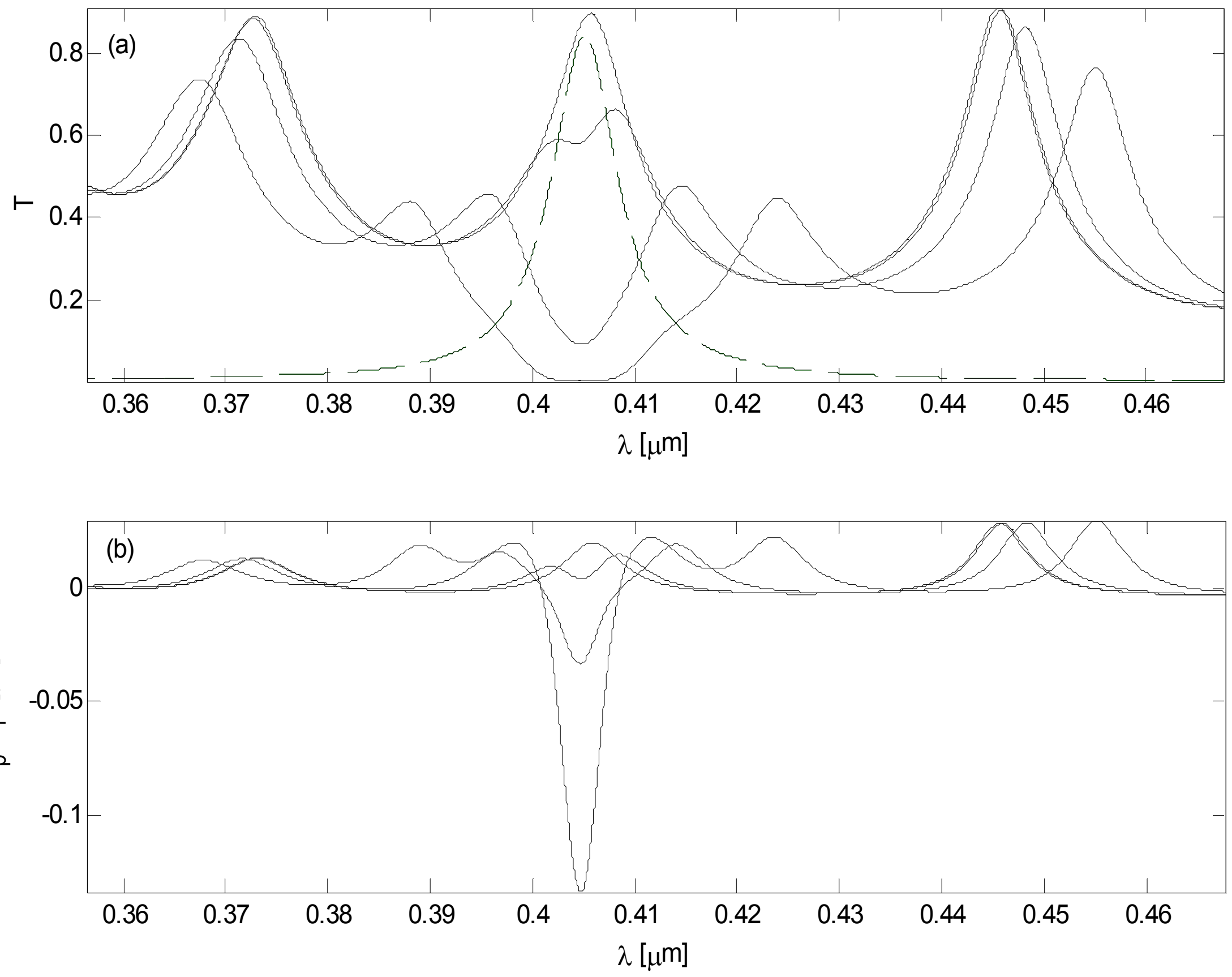

Fig.3 M Kulkarni..,"Control of superluminal …."

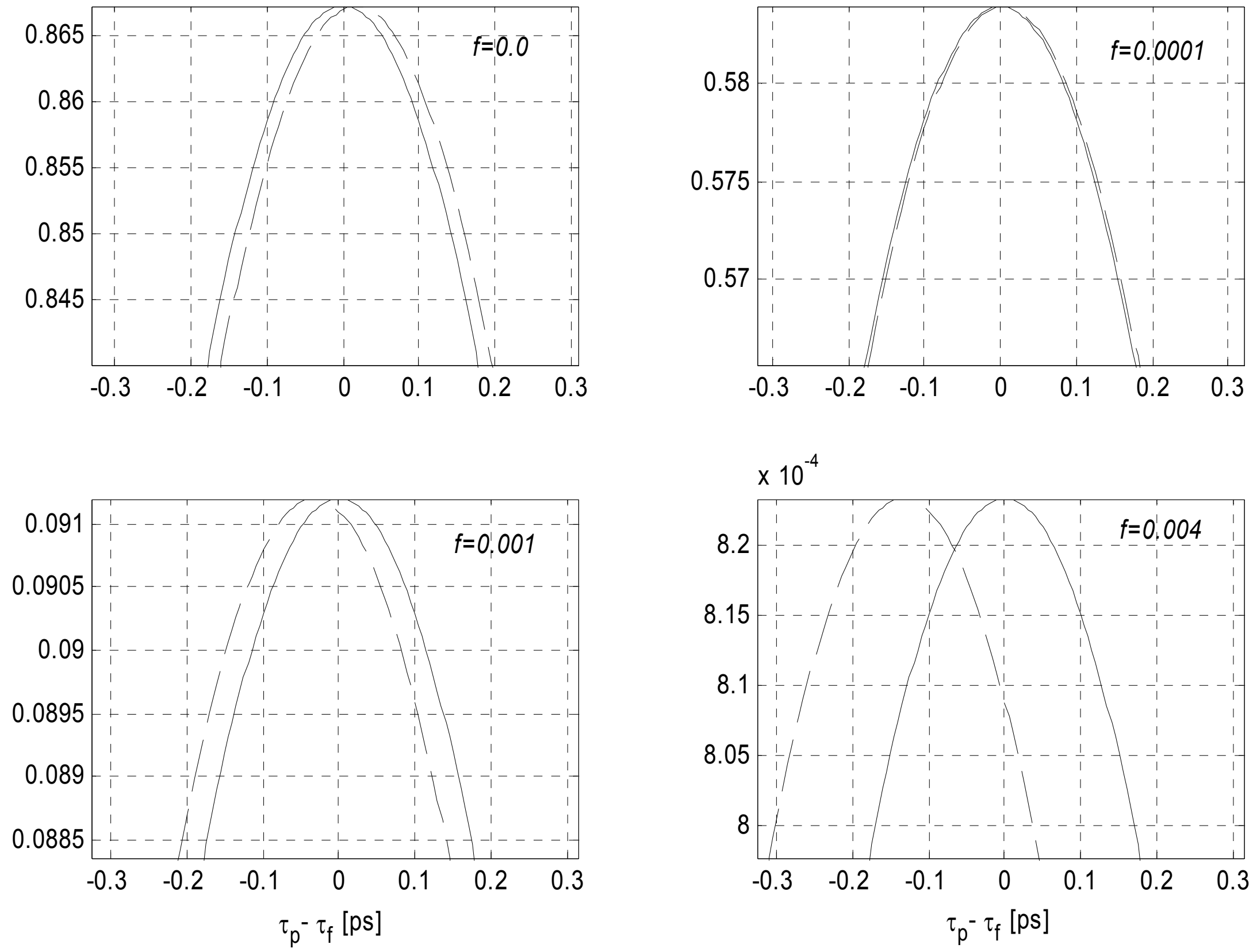

Fig.4 M Kulkarni..,"Control of superluminal …."

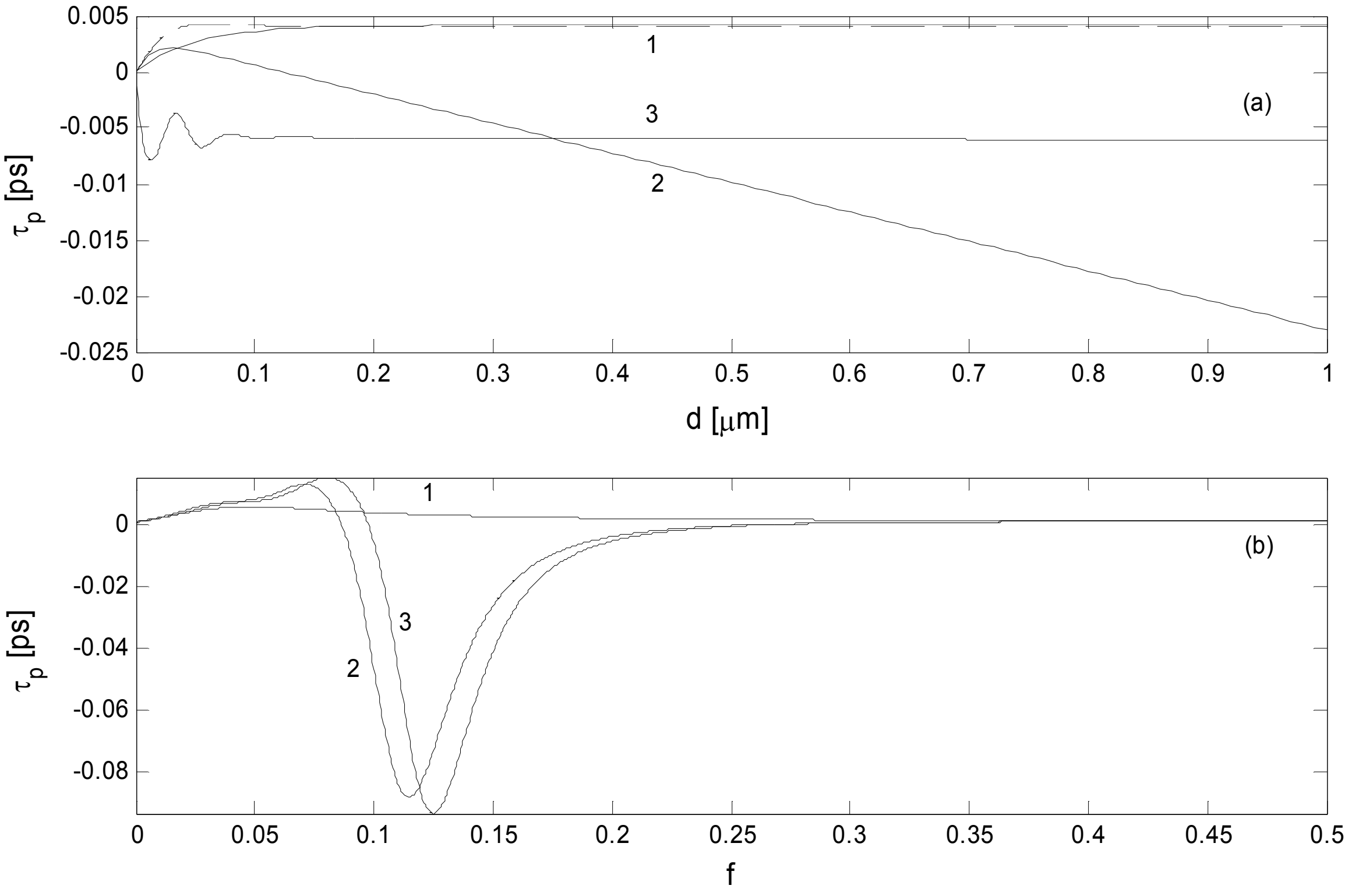

Fig.5 M Kulkarni..,"Control of superluminal …."

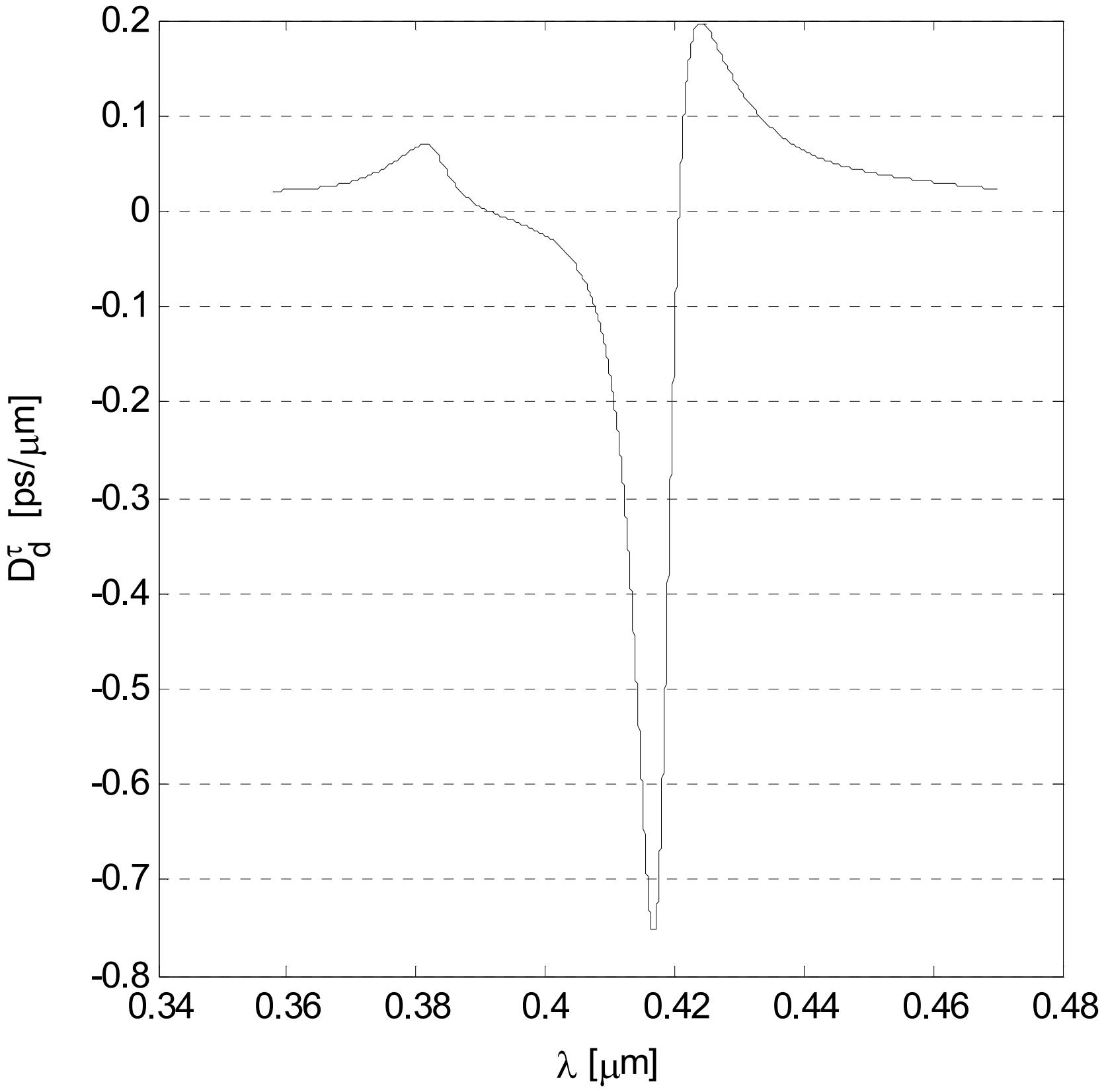

Inset to Fig.5a M Kulkarni..,"Control of superluminal …."